\documentclass[]{spie}  

\usepackage{amsmath,amsfonts,amssymb}
\usepackage{graphicx}
\usepackage[colorlinks=true, allcolors=blue]{hyperref}
\usepackage{siunitx}
\newcommand{\SIadj}[2]{\SI[number-unit-product={\text{-}}]{#1}{#2}}

\usepackage{subcaption}
\title{Physical Optics Analysis of Polarization Effects in SO LAT Reflectors}

\author[a]{Xiaodong Ren}
\author[b]{Patricio A. Gallardo}
\author[a]{Jon E. Gudmundsson}

\affil[a]{Science Institute, University of Iceland, 107 Reykjavik, Iceland}
\affil[b]{Department of Physics and Astronomy, University of Pennsylvania, Philadelphia, PA, USA}

\authorinfo{Further author information: (Send correspondence to X. Ren)\\ E-mail: xren@hi.is}

\begin{document} 
\maketitle

\begin{abstract}
We present a physical-optics analysis of the polarization response of the reflective optical of the Simons Observatory Large Aperture Telescope (SO LAT). Far-field co-polar and cross-polar beam patterns are simulated for representative feedhorn positions, and the corresponding polarization-angle offsets are evaluated. The results show that the crossed-Dragone reflector system has good polarization performance. Off-axis feed positions introduce small, position-dependent offsets in the far-field polarization orientation, with a maximum value of $0.23^\circ$, due to optical asymmetry. The relative polarization-angle offset under feedhorn polarization rotation, however, remains consistent with zero. No significant frequency dependence is found over the frequencies considered. These results provide a practical reference for polarization-angle calibration and for assessing reflector-induced polarization systematics in the SO LAT.
\end{abstract}


\section{INTRODUCTION}
\label{sec:intro}  
Measurements of the polarization anisotropies of the cosmic microwave background (CMB) are a major objective of current and next-generation observational cosmology experiments. CMB polarization encodes information about primordial density perturbations, the reionization history of the Universe, and possible primordial gravitational waves generated during inflation \cite{SOScience_2019,Minami2020}. Since the polarized signal is much weaker than the temperature anisotropy, accurate characterization of instrumental systematic effects is essential. Optical systems can alter the polarization state of incoming radiation through polarization rotation, cross-polarization response, and instrumental polarization, producing spurious signals that may contaminate the measured polarization anisotropies \cite{o2007systematic,PhysRevD.77.083003}. Therefore, accurately measuring and calibrating polarization rotation induced by the optical system is an important step toward accurate CMB polarization observations \cite{cornelison2022improved}. It is also necessary to quantify systematic polarization errors induced by telescope optics, as such analysis can provide useful references for polarization calibration and guidance for the optical design of CMB telescopes\cite{franco2003systematic}.

The Simons Observatory Large Aperture Telescope (SO LAT) is a 6 m coma-corrected crossed-Dragone (CD) telescope designed to provide high angular resolution and a large field of view for CMB observations\cite{parshley2018ccat, Dicker2018, Gallardo2018, Gudmundsson:21, Xu2021}. Its optical system consists of two large off-axis, panel-segmented reflectors followed by the cryogenic Large Aperture Telescope Receiver (LATR), which houses 13 optics tubes~\cite{Zhu_2021,2025ApJS27934B}. Each optics tube contains cold refractive optics composed of three silicon lenses that re-image the telescope focal plane onto the detector arrays. 
The reflective optics offers the advantages of zero aperture blockage and a large field of view while maintaining low cross-polarization. Additional coma-correction terms are applied to the two reflectors to further improve the field of view, although this comes at the cost of a more curved focal plane\cite{parshley2018optical}. Nevertheless, residual polarization effects from the reflector system must still be characterized, especially for off-axis detectors and wide-field observations.

In this paper, we investigate reflector-induced polarization effects from the SO LAT crossed-Dragone reflectors across the focal plane. The analysis is restricted to the crossed-Dragone primary (M1) and secondary (M2) reflectors, while the cryogenic re-imaging optics, including the silicon lenses, filters, Lyot stop, and detector coupling, are not included. The results should therefore be interpreted as a reflector-only polarization characterization, rather than a complete end-to-end polarization systematic budget for the SO LAT. The cold refractive optics of the SO LAT will be analyzed in future work. We expect that this reflector-only study provides a simulation-based reference for future end-to-end optical modeling and polarization angle calibration of the SO LAT.

To carry out the polarization analysis for the reflectors, we use a physical optics (PO) method\cite{collin1969antenna} to simulate the far-field polarized beam patterns of the reflectors across the focal plane. The simulations are performed at 90, 150 and \SI{230}{\giga\hertz}.  Both the commercial software TICRA Tools (also known as GRASP) and a two-step PO analysis method, originally developed for the Fred Young Sub-millimeter Telescope (FYST) holography experiment\cite{xd_holo_FYST}, are employed to compute polarized beam patterns on the sky for different receiver positions in the focal plane.

\section{Far-field Beam Simulations}
\subsection{Overview}
\begin{figure}[h]
    \centering
    \includegraphics[width=1\linewidth]{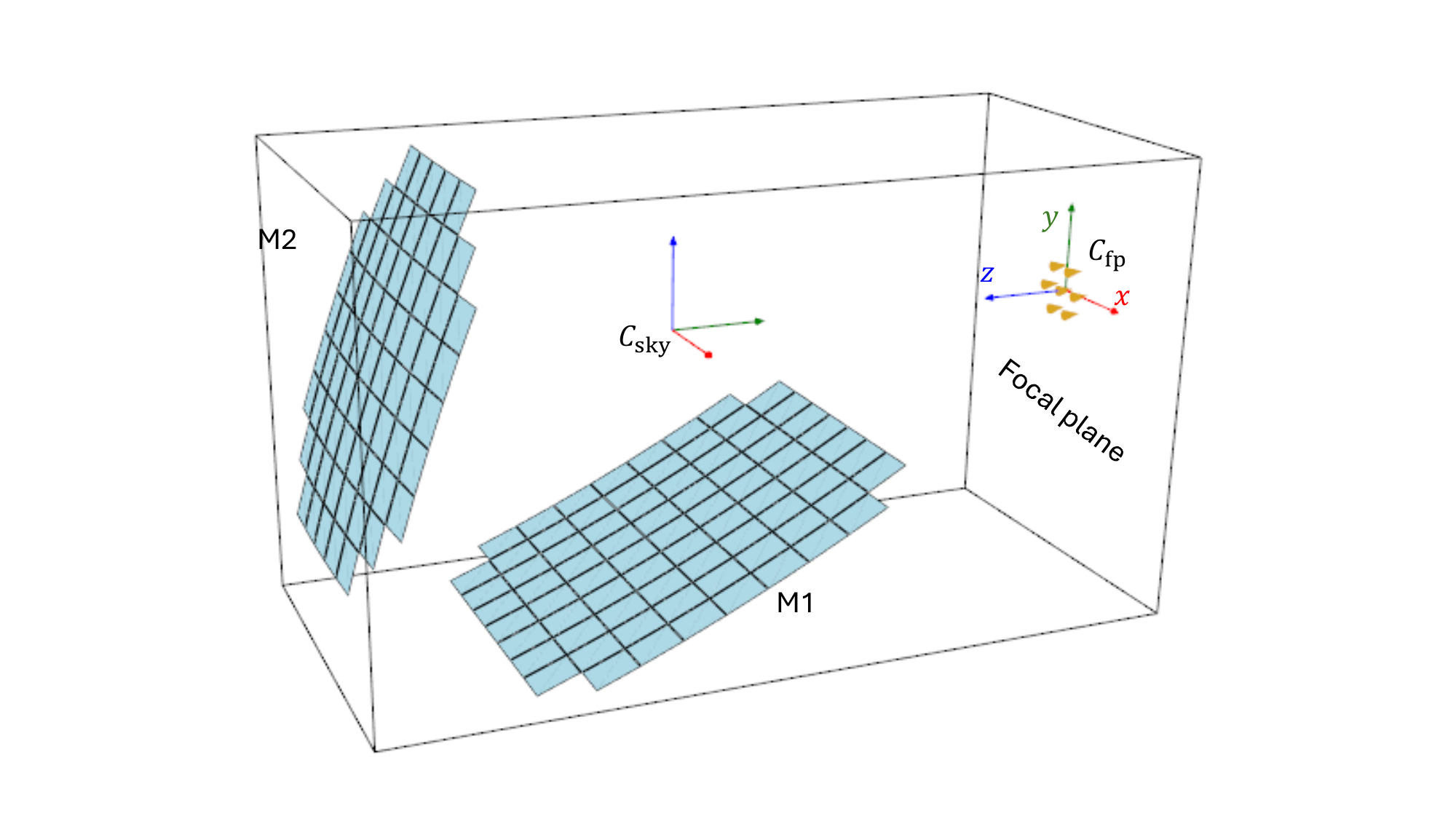}
    \caption{Schematic of the SO LAT optical layout with feedhorns placed on a flat focal plane.}
    \label{fig:SOLAT_layout}
\end{figure}

The SO LAT consists of two similarly sized reflectors, M1 and M2, which are segmented into 146 rectangular panels in total. The nominal inter-panel gap is \SI{1.2}{\milli\meter} at room temperature (\SI{300}{\kelvin}). This gap size is designed to accommodate thermal expansion at higher temperatures, such as during transport near the equator, and to prevent adjacent panels from coming into contact and deforming. Figure \ref{fig:SOLAT_layout} shows a schematic of the optical layout. In the simulation configuration, the feedhorn is placed at selected positions across the \SIadj{2}{\meter} diameter focal plane. An ideal linearly polarized Gaussian beam is adopted as the feedhorn radiation pattern to illuminate the reflectors, and its beam size is controlled by the illumination edge taper. In the practical observation case, signals from the distant sky are collected by the primary reflector, reflected by the secondary reflector and then coupled to the detectors in the focal plane. In the simulations presented in this paper, it is more convenient to simulate the equivalent time-reversed process, based on the antenna reciprocity theorem. In this configuration, the signal starts from the feedhorn, is scattered by the two reflectors, and propagates toward the sky. The calculated far-field beam is used to characterize the polarization response of the SO LAT reflector system.
\subsection{Simulation methods}\label{subsection:po}
The polarization response of the SO LAT reflectors is studied by simulating their far-field beam patterns using PO analysis. The commercial software TICRA Tools is used to perform the analysis. Compared with geometrical optics (GO), which describes the optical system in the ray limit and neglects wavelength-dependent effects, PO analysis retains the wave nature of electromagnetic propagation and is therefore suitable for predicting the far-field radiation patterns of reflector antennas, including diffraction effects, the cross-polarization response, and the spatial distribution of polarization properties across the main-beam region. Since the reflectors are panel-segmented, the PO approximation only captures the dominant reflection from the panel surfaces but does not fully describe the discontinuities at panel edges. In practice, these discontinuities generate additional fringe currents near the edges, over a spatial scale of order the wavelength. For a fixed panel geometry, the relative importance of these edge-diffraction effects is therefore expected to decrease at shorter wavelengths. To evaluate this contribution, the physical theory of diffraction\cite{ufimtsev2014fundamentals} (PTD) technique in GRASP is employed to analyze the panel edges effects. 

For the systematic focal plane study, we also use a two-step PO analysis method, which is originally developed for the Fred Young Submillimeter Telescope (FYST) holography experiment\cite{xd_holo_FYST}. Since the full PO simulations in GRASP are computationally intensive, because of the two large and closely-spaced reflectors. The total computational cost increases substantially when the analysis requires repeated simulations over multiple receiver positions, feed polarization angles and multiple observing frequencies. The two-step PO method provides a faster way to compute the reflector far-field beams across the focal plane by introducing an auxiliary plane at the intermediate focal plane of M2, located approximately \SI{30}{\meter} behind the secondary reflector. With this auxiliary plane, the propagation from M2 to M1 is divided into two focused calculation steps, avoiding the need to sample a highly defocused field directly over the full M1 surface. This greatly reduces the required number of sampling points on the reflector panels and therefore lowers the computational cost. The accuracy of this method has been validated against GRASP simulations, and the field accuracy in the main beam region is better than \SI{-60}{\decibel} relative to the beam peak. Therefore, this two-step method is used for the large parameter study presented in this paper.
 
\subsection{Coordinate frames and polarization conventions}\label{subsection:method_polar}
Before starting the polarization response analysis, it is necessary to define the reference frames used for specifying the feedhorn position, its polarization orientation, and the corresponding beam and its polarization orientation on the sky. The receiver position is specified in the focal plane coordinate system, $C_{\rm{fp}}$, as shown in Figure\,\ref{fig:SOLAT_layout}. The feedhorn polarization angle, denoted by $\varphi_{\rm feed}$, is defined with respect to the $x$-axis. The simulated far-field beam, $\mathbf{E}_{\rm far}$, is described in the sky coordinate system, $C_{\rm{sky}}$, as indicated in Figure \ref{fig:SOLAT_layout}. The $z$-axis of this coordinate system points toward the telescope boresight, and the far-field beam is represented using the $(u,v)$ coordinate. Here, $u$ and $v$ are the far-field direction-cosine coordinates, defined as
\begin{equation}
    u=\sin\theta\cos\phi,\qquad v=\sin\theta\sin\phi,
\end{equation}
where $\theta$ is the polar angle measured from the boresight direction and $\phi$ is the azimuthal angle. Using Ludwig's third definition\cite{ludwig1973}, the vector radiation field can be decomposed into its co-polar ($\mathbf{E}_{\rm co}$) and cross-polar ($\mathbf{E}_{\rm cx}$) components. 

Based on this convention, two methods are used to evaluate the polarization angle of the far-field beam. One approach (\textbf{Method 1}) is to calculate the beam-averaged polarization angle from the Stokes $Q$ and $U$ beam maps by integrating over the main beam region. In this work, the main beam region is defined as the region where the gain is above $-20$ dB relative to the beam peak:
\begin{equation}
    \varphi_{\rm beam}= \frac{1}{2}\mathrm{atan2}
    \left(
    \int_{\Omega_{\rm MB}} U(u,v)\,d\Omega\, ,
    \int_{\Omega_{\rm MB}} Q(u,v)\,d\Omega
    \right),
    \label{eq:polar_method1}
\end{equation}
where $\Omega_{\rm MB}$ denotes the main-beam region.

Alternatively, in the second approach (\textbf{Method 2}) the polarization angle can be determined by rotating the reference polarization basis and finding the rotation angle, $\varphi$, that maximizes the ratio between the integrated co-polar and cross-polar power over the main beam region. This ratio is referred to as the cross-polarization discrimination\cite{franco2003systematic} (XPD) of the optical system:
\begin{equation}
    {\rm XPD}(\varphi) = 
    \frac{
    \int_{\Omega_{\rm MB}} |\mathbf{E}^{\varphi}_{\rm co}(u,v)|^2 d\Omega
    }{
    \int_{\Omega_{\rm MB}} |\mathbf{E}^{\varphi}_{\rm cx}(u,v)|^2 d\Omega
    },
    \label{eq:polar_method2}
\end{equation}
where $\mathbf{E}^{\varphi}_{\rm co}$ and $\mathbf{E}^{\varphi}_{\rm cx}$ are the co-polar and cross-polar fields after rotating the reference frame by angle of $\varphi$. Therefore the beam polarization angle is then defined as
\begin{equation}
    \varphi_{\rm beam} = \arg\max_{\varphi}{\rm XPD}(\varphi).
\end{equation}

With the above definitions, the polarization orientation of the telescope far-field beam can be determined with respect to the adopted optical reference frame. We also evaluate the relative polarization-angle response when the feedhorn polarization is rotated from an initial angle, $\varphi^{i}_{\rm feed}$, to a target angle, $\varphi^{t}_{\rm feed}$, and the corresponding far-field beam polarization angle changes from $\varphi^{i}_{\rm beam}$ to $\varphi^{t}_{\rm beam}$. For an ideal optical system, the beam polarization angle should rotate by the same amount as the feedhorn polarization angle. However, for an off-axis reflecting system, especially for feedhorns displaced from the optical axis, these two rotation angles may differ. We define this difference as the relative polarization-angle offset,
\begin{equation}
\Delta \varphi_{\rm rel} = \left(\varphi^{t}_{\rm beam}-\varphi^{i}_{\rm beam}\right) - \left(\varphi^{t}_{\rm feed}-\varphi^{i}_{\rm feed}\right).
\end{equation}
These definitions are used below to quantify both the far-field polarization orientation with respect to the defined optical reference frame and the relative polarization-angle offset introduced by the LAT reflector system.

\section{Results}
In this section, we present the far field polarization response of the SO LAT reflector system, computed using the beam simulation methods described in Sect.~\ref{subsection:po}. We compute the far field beam patterns for feedhorns placed at different positions in the focal plane, from near the optical axis to the edge of the field of view. The feedhorn illumination is modeled as an ideal circular Gaussian beam, with the beam width set by the prescribed telescope-aperture edge taper ranging from $-8$ dB to $-20$ dB. For each feed position, we study the telescope beam polarization as a function of the feedhorn polarization angle, $\varphi_{\rm beam}$, which is varied from \SI{0}{\degree} to \SI{180}{\degree} in steps of \SI{15}{\degree}. The resulting beam polarization angle is evaluated using the two methods described in Sect.~\ref{subsection:method_polar}. The simulations are performed at three frequencies, 90, 150 and \SI{230}{\giga\hertz}. This allows us to investigate how feed position, observing frequency, feed illumination, and off-axis reflection affect the preservation of the polarization orientation through the telescope. Although this work focuses on the SO LAT reflectors, the results may also be relevant to other mm-wave polarimetric experiments employing similar off-axis or crossed-Dragone optical systems. 

Before discussing the polarization-angle response in detail, we first show representative simulated beam patterns and compare the GRASP PO results with those obtained from the two-step PO analysis. We then assess the impact of panel-edge diffraction on the beam polarization properties.

\subsection{Far-field beam simulations and effect of reflector segmentation}
\begin{figure}[h]
    \centering
    \includegraphics[width=0.65\linewidth]{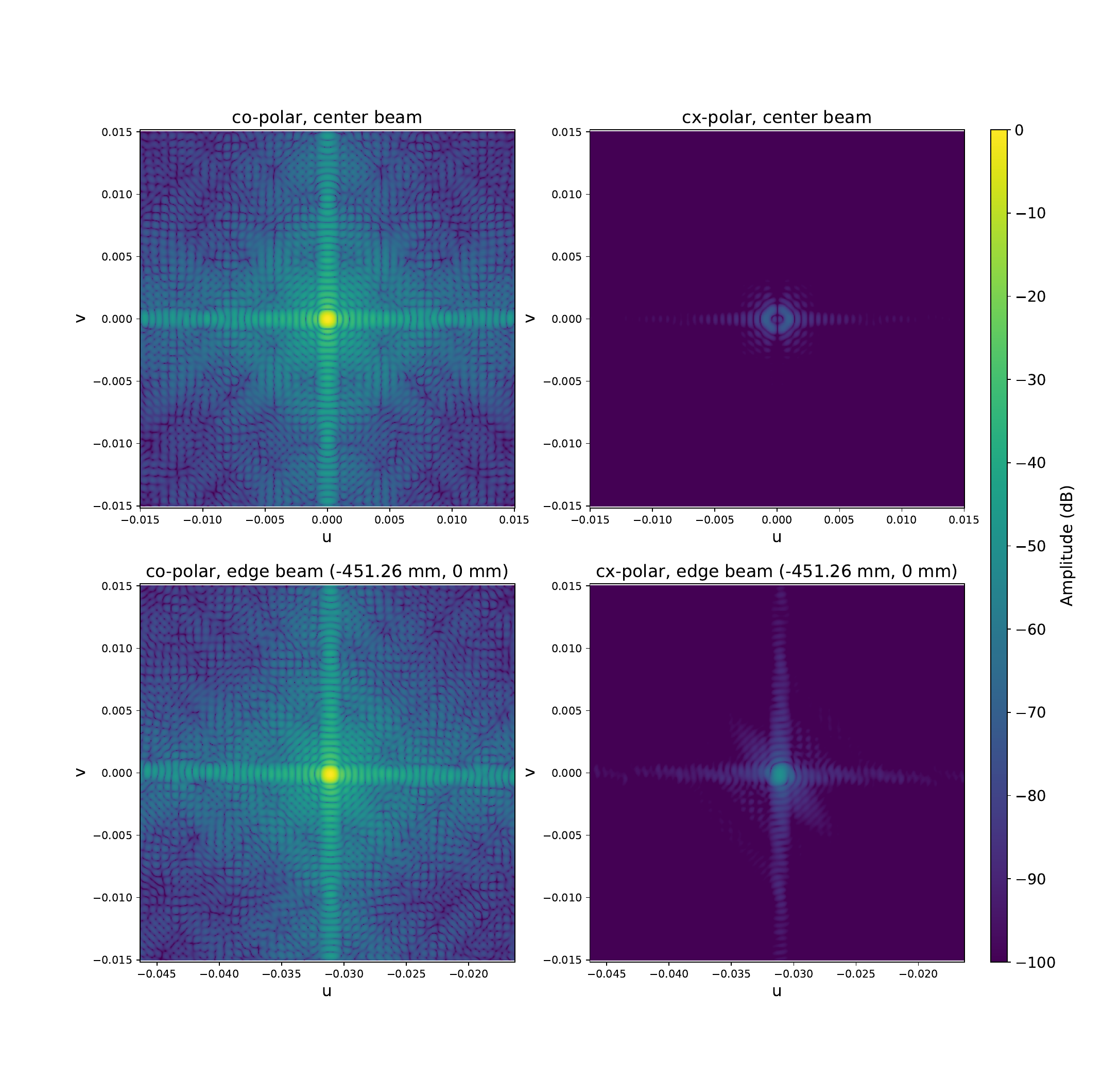}
    \caption{Co- and cross-polarization beam patterns of the SO LAT reflectors for a Gaussian feedhorn operating at 90 GHz and located at (\SI{-451.26}{\milli\meter}, \SI{0}{\milli\meter}) in the focal plane, corresponding to an angular distance of approximately $\SI{1.8}{\degree}$ from the boresight on the sky.}
    \label{fig:90GHz_edge_center_beams}
\end{figure}

\begin{figure}[h]
    \centering
    \includegraphics[width=0.5\linewidth]{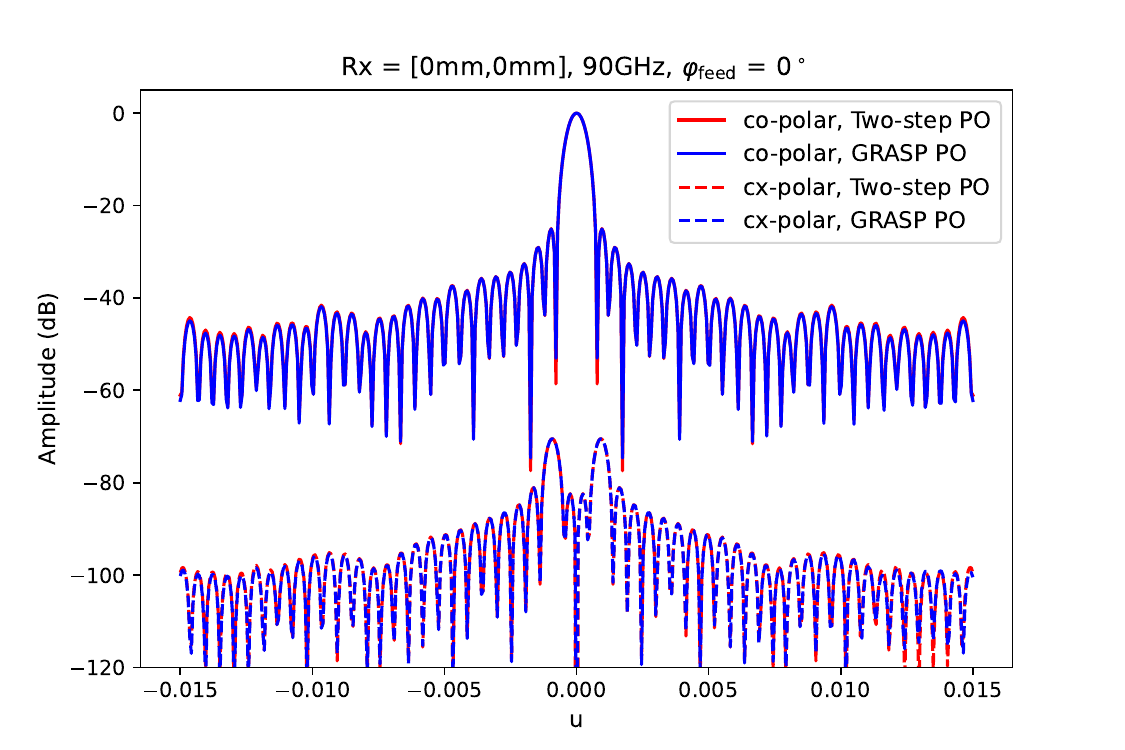}
    \caption{Normalized 90 GHz beam patterns at the focal plane center of the SO LAT. Comparison between GRASP reference data and the two-step PO method for both co-polar/cross-polar components.}
    \label{fig:cutplots_comparison}
\end{figure}

We first validate the two-step PO analysis by comparing the resulting far-field beam patterns with those obtained from full GRASP PO simulations. The comparison is performed for two representative feedhorn positions in the focal plane: one at the focal plane center and one at an off-axis focal plane position, located at (\SI{451.26}{\milli\meter}, \SI{0}{\milli\meter}) in the $C_{\rm fp}$ coordinate system. Figure~\ref{fig:90GHz_edge_center_beams} shows the 90 GHz beam maps of the reflectors for the feedhorn located at center and edge of the focal plane. In the simulations the feedhorn beam is chosen to produce a \SI{-12}{\deci\bel} illumination edge taper. 
Figure~\ref{fig:cutplots_comparison} compares the co-polar and cross-polar beam cuts along the $u$-direction, obtained from the full GRASP PO simulations and from the two-step PO analysis. For both the co-polar and cross-polar components, the two sets of beam cuts show good agreement.

\begin{figure}[h]
    \centering
    \includegraphics[width=0.5\linewidth]{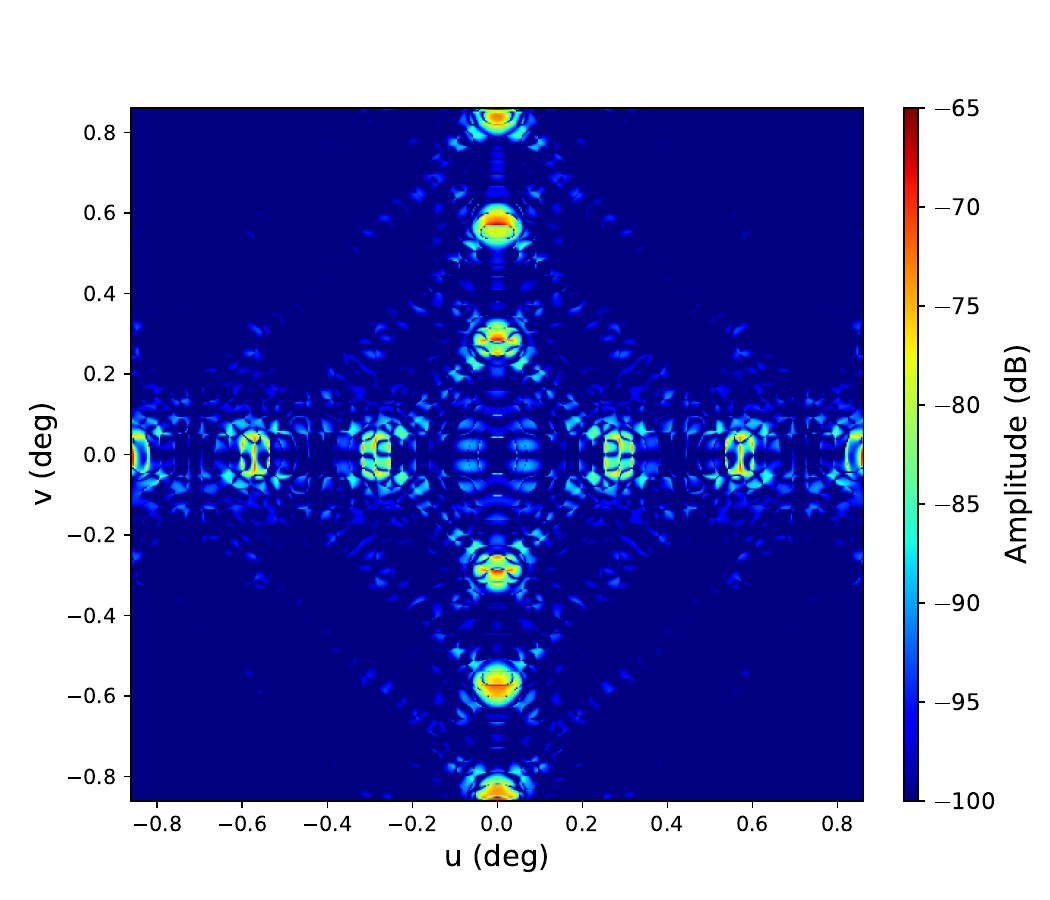}
    \caption{Simulated \SIadj{90}{\giga\hertz} diffraction pattern generated by the panel edges of the segmented reflectors for the case of feedhorn polarized along $x$-axis ($\varphi_{\rm feed}=0^{\circ}$). For the small angular range shown here, the $u$ and $v$ coordinates are expressed in degrees for readability.}
    \label{fig:panel_edge_diffraction}
\end{figure}

The effect of reflector segmentation is assessed by comparing GRASP PO-only simulations with PO+PTD simulations, where the PTD analysis accounts for diffraction from the panel edges. Figure~\ref{fig:panel_edge_diffraction} shows the difference between the PO+PTD and PO-only simulations for an input feedhorn beam linearly polarized at \SI{0}{\degree}, corresponding to polarization along the $x$-axis. It can be seen that the panel-edge diffraction contribution remains below approximately \SI{-65}{\deci\bel} relative to the main-beam peak. Therefore, we assume that the effect has a negligible impact on the beam polarization properties considered in this work.

\subsection{Polarization-angle response of off-axis beams}
\label{section:polar_position_polar}
\begin{figure}[h]
    \centering
    \includegraphics[width=0.5\linewidth]{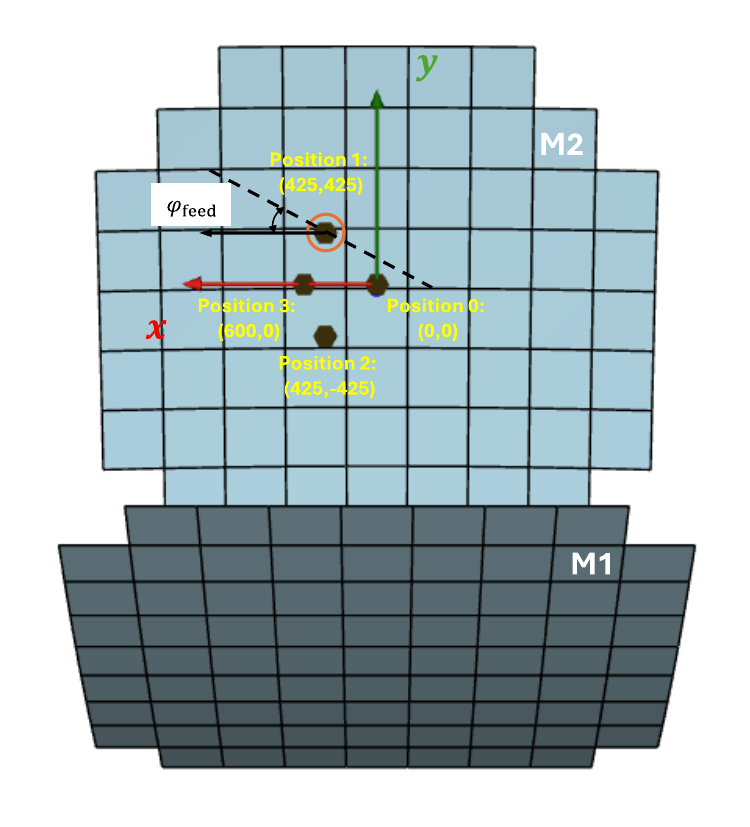}
    \caption{Focal plane layout used in the far field beam simulations and polarization analysis. The layout is shown in a perspective view from the focal plane looking toward M2. The selected feedhorn positions are labeled as Position 0 and Positions 1--3, with their corresponding focal plane coordinates indicated in millimeters.}
    \label{fig:fp_config}
\end{figure}

Figure~\ref{fig:fp_config} shows the center point (position 0) and the three off-axis feed positions (position 1-3) selected for polarization performance analysis. In the focal plane coordinate system, $C_{\rm fp}$, Position 1 is located at $(425,425)$ mm, Position 2 at $(425,-425)$ mm, and Position 3 at $(600,0)$ mm. For each position, the input feedhorn polarization angle $\varphi_{\rm feed}$ is rotated from $0^\circ$ to $180^\circ$ in steps of $15^\circ$ to evaluate the corresponding beam polarization-angle response. Only one side of the focal plane is shown here, since the receiver positions on the opposite side are related by symmetry to the selected positions. The corresponding polarization-orientation offsets are therefore expected to have the same magnitude but opposite sign. This symmetry behavior, as well as the order of magnitude of the resulting polarization-angle offsets, is consistent with the optical analysis reported by Murphy et al.~\cite{Murphy:24}. In the simulations, the illumination edge taper is still set to \SI{-12}{\decibel} and the study is carried out for 90, 150 and \SI{230}{\giga\hertz}. 
\begin{table}[t]
\caption{Results of the polarization angle of the far field beam of SO LAT reflectors at \SI{90}{\giga\hertz} using Method 2. The illumination edge taper is \SI{-12}{\decibel}.} 
\label{tab:polarization_offset_90GHz}
        \begin{subtable}[b]{\textwidth}
            \begin{center}  
                \begin{tabular}{|c|c|c|c|c|c|c|c|c|c|c|c|c|}
                    \hline
                     \rule[-1ex]{0pt}{3.5ex} 90 GHz 
                     & \multicolumn{3}{c|}{Position 1} 
                     &  \multicolumn{3}{c|}{Position 2} 
                     &  \multicolumn{3}{c|}{Position 3}  \\
                    \hline
                    \rule[-1ex]{0pt}{3.5ex}  $\varphi_{\rm feed}$ 
                    & $\varphi_{\rm beam}$ & $\Delta\varphi_{\rm rel}$ & $\rm{XPD}_{\rm max}$ 
                    & $\varphi_{\rm beam}$ & $\Delta\varphi_{\rm rel}$ & $\rm{XPD}_{\rm max}$
                    & $\varphi_{\rm beam}$ & $\Delta\varphi_{\rm rel}$ & $\rm{XPD}_{\rm max}$  \\
                    \rule[-1ex]{0pt}{3.5ex} 
                    & ($^\circ$) & ($^\circ$) & (dB)
                    & ($^\circ$) & ($^\circ$) & (dB)
                    & ($^\circ$) & ($^\circ$) & (dB)  \\
                    \hline 
                    \rule[-1ex]{0pt}{3.5ex} $0^{\circ}$ 
                    & 0.17 & 0.0 & 56.78
                    & 0.16 & 0.0 & 56.27
                    & 0.23 & 0.0 & 58.75 \\
                    \rule[-1ex]{0pt}{3.5ex} $ 15^{\circ}$
                    & 15.17 & 0.0 & 56.78
                    & 15.16 & 0.0 & 56.27
                    & 15.23 & 0.0 & 58.74 \\
                    \rule[-1ex]{0pt}{3.5ex} $ 30^{\circ}$
                    & 30.17 & 0.0 & 56.78
                    & 30.16 & 0.0 & 56.26
                    & 30.23 & 0.0 & 58.73 \\
                    \rule[-1ex]{0pt}{3.5ex} $ 45^{\circ}$
                    & 45.17 & 0.0 & 56.77
                    & 45.16 & 0.0 & 56.26
                    & 45.23 & 0.0 & 58.72 \\
                    \rule[-1ex]{0pt}{3.5ex} $ 60^{\circ}$
                    & 60.17 & 0.0 & 56.77
                    & 60.16 & 0.0 & 56.27
                    & 60.23 & 0.0 & 58.72 \\
                    \rule[-1ex]{0pt}{3.5ex} $ 75^{\circ}$
                    & 75.17 & 0.0 & 56.77
                    & 75.16 & 0.0 & 56.27
                    & 75.23 & 0.0 & 58.71 \\
                    \rule[-1ex]{0pt}{3.5ex} $ 90^{\circ}$
                    & 90.17 & 0.0 & 56.76
                    & 90.16 & 0.0 & 56.28
                    & 90.23 & 0.0 & 58.71 \\
                    \rule[-1ex]{0pt}{3.5ex} $ 105^{\circ}$
                    & 105.17 & 0.0 & 56.76
                    & 105.16 & 0.0 & 56.29
                    & 105.23 & 0.0 & 58.72 \\
                    \rule[-1ex]{0pt}{3.5ex} $ 120^{\circ}$
                    & 120.17 & 0.0 & 56.77
                    & 120.16 & 0.0 & 56.29
                    & 120.23 & 0.0 & 58.73 \\
                    \rule[-1ex]{0pt}{3.5ex} $ 135^{\circ}$
                    & 135.17 & 0.0 & 56.77
                    & 135.16 & 0.0 & 56.29
                    & 135.23 & 0.0 & 58.74 \\
                    \rule[-1ex]{0pt}{3.5ex} $ 150^{\circ}$
                    & 150.17 & 0.0 & 56.78
                    & 150.16 & 0.0 & 56.29
                    & 150.23 & 0.0 & 58.74 \\
                    \rule[-1ex]{0pt}{3.5ex} $ 165^{\circ}$
                    & 165.17 & 0.0 & 56.78
                    & 165.16 & 0.0 & 56.28
                    & 165.23 & 0.0 & 58.75 \\
                    \rule[-1ex]{0pt}{3.5ex} $ 180^{\circ}$
                    & 180.17 & 0.0 & 56.78
                    & 180.16 & 0.0 & 56.27
                    & 180.23 & 0.0 & 58.75 \\
                    \hline
                \end{tabular}
            \end{center}
        \end{subtable}     
\end{table} 
\\

\begin{table}[ht]
\caption{Results of the polarization angle of the far-field beam of SO LAT reflectors at 150 GHz using Method 2. The illumination edge taper is \SI{-12}{\decibel}.} 
\label{tab:polarization_offset_150GHz}
        \begin{subtable}[b]{\textwidth}
            \begin{center}  
                \begin{tabular}{|c|c|c|c|c|c|c|c|c|c|c|c|c|}
                    \hline
                     \rule[-1ex]{0pt}{3.5ex} 150 GHz 
                     & \multicolumn{3}{c|}{Position 1} 
                     &  \multicolumn{3}{c|}{Position 2} 
                     &  \multicolumn{3}{c|}{Position 3}  \\
                    \hline
                    \rule[-1ex]{0pt}{3.5ex}  $\varphi_{\rm feed}$ 
                    & $\varphi_{\rm beam}$ & $\Delta\varphi_{\rm rel}$ & $\rm{XPD}_{\rm max}$ 
                    & $\varphi_{\rm beam}$ & $\Delta\varphi_{\rm rel}$ & $\rm{XPD}_{\rm max}$
                    & $\varphi_{\rm beam}$ & $\Delta\varphi_{\rm rel}$ & $\rm{XPD}_{\rm max}$  \\
                    \rule[-1ex]{0pt}{3.5ex} 
                    & ($^\circ$) & ($^\circ$) & (dB)
                    & ($^\circ$) & ($^\circ$) & (dB)
                    & ($^\circ$) & ($^\circ$) & (dB)  \\
                    \hline 
                    \rule[-1ex]{0pt}{3.5ex} $0^{\circ}$ 
                    & 0.17 & 0.0 & 56.73
                    & 0.16 & 0.0 & 56.13
                    & 0.23 & 0.0 & 58.58 \\
                    \rule[-1ex]{0pt}{3.5ex} $ 15 ^{\circ}$
                    & 15.17 & 0.0 & 56.73
                    & 15.16 & 0.0 & 56.12
                    & 15.23 & 0.0 & 58.57 \\
                    \rule[-1ex]{0pt}{3.5ex} $ 30 ^{\circ}$
                    & 30.17 & 0.0 & 56.73
                    & 30.16 & 0.0 & 56.12
                    & 30.23 & 0.0 & 58.56 \\
                    \rule[-1ex]{0pt}{3.5ex} $ 45 ^{\circ}$
                    & 45.17 & 0.0 & 56.73
                    & 45.16 & 0.0 & 56.12
                    & 45.23 & 0.0 & 58.56 \\
                    \rule[-1ex]{0pt}{3.5ex} $ 60 ^{\circ}$
                    & 60.17 & 0.0 & 56.73
                    & 60.16 & 0.0 & 56.12
                    & 60.23 & 0.0 & 58.55 \\
                    \rule[-1ex]{0pt}{3.5ex} $ 75 ^{\circ}$
                    & 75.17 & 0.0 & 56.73
                    & 75.16 & 0.0 & 56.13
                    & 75.23 & 0.0 & 58.54 \\
                    \rule[-1ex]{0pt}{3.5ex} $ 90 ^{\circ}$
                    & 90.17 & 0.0 & 56.73
                    & 90.16 & 0.0 & 56.13
                    & 90.23 & 0.0 & 58.54 \\
                    \rule[-1ex]{0pt}{3.5ex} $ 105 ^{\circ}$
                    & 105.17 & 0.0 & 56.73
                    & 105.16 & 0.0 & 56.14
                    & 105.23 & 0.0 & 58.54 \\
                    \rule[-1ex]{0pt}{3.5ex} $ 120 ^{\circ}$
                    & 120.17 & 0.0 & 56.73
                    & 120.16 & 0.0 & 56.14
                    & 120.23 & 0.0 & 58.55 \\
                    \rule[-1ex]{0pt}{3.5ex} $ 135 ^{\circ}$
                    & 135.17 & 0.0 & 56.73
                    & 135.16 & 0.0 & 56.14
                    & 135.23 & 0.0 & 58.56 \\
                    \rule[-1ex]{0pt}{3.5ex} $ 150 ^{\circ}$
                    & 150.17 & 0.0 & 56.73
                    & 150.16 & 0.0 & 56.14
                    & 150.23 & 0.0 & 58.57 \\
                    \rule[-1ex]{0pt}{3.5ex} $ 165 ^{\circ}$
                    & 165.17 & 0.0 & 56.73
                    & 165.16 & 0.0 & 56.13
                    & 165.23 & 0.0 & 58.57 \\
                    \rule[-1ex]{0pt}{3.5ex} $ 180 ^{\circ}$
                    & 180.17 & 0.0 & 56.73
                    & 180.16 & 0.0 & 56.13
                    & 180.23 & 0.0 & 58.58 \\
                    \hline
                \end{tabular}
            \end{center}
        \end{subtable}     
\end{table}

\begin{table}[ht]
\caption{Results of the polarization angle of the far-field beam of SO LAT reflectors at 230 GHz using Method 2. The illumination edge taper is \SI{-12}{\decibel}.} 
\label{tab:polarization_offset_230GHz}
        \begin{subtable}[b]{\textwidth}
            \begin{center}  
                \begin{tabular}{|c|c|c|c|c|c|c|c|c|c|c|c|c|}
                    \hline
                     \rule[-1ex]{0pt}{3.5ex} 230 GHz 
                     & \multicolumn{3}{c|}{Position 1} 
                     &  \multicolumn{3}{c|}{Position 2} 
                     &  \multicolumn{3}{c|}{Position 3}  \\
                    \hline
                    \rule[-1ex]{0pt}{3.5ex}  $\varphi_{\rm feed}$ 
                    & $\varphi_{\rm beam}$ & $\Delta\varphi_{\rm rel}$ & $\rm{XPD}_{\rm max}$ 
                    & $\varphi_{\rm beam}$ & $\Delta\varphi_{\rm rel}$ & $\rm{XPD}_{\rm max}$
                    & $\varphi_{\rm beam}$ & $\Delta\varphi_{\rm rel}$ & $\rm{XPD}_{\rm max}$  \\
                    \rule[-1ex]{0pt}{3.5ex} 
                    & ($^\circ$) & ($^\circ$) & (dB)
                    & ($^\circ$) & ($^\circ$) & (dB)
                    & ($^\circ$) & ($^\circ$) & (dB)  \\
                    \hline 
                    \rule[-1ex]{0pt}{3.5ex} $0^{\circ}$ 
                    & 0.17 & 0.0 & 56.73
                    & 0.16 & 0.0 & 55.95
                    & 0.23 & 0.0 & 58.46 \\
                    \rule[-1ex]{0pt}{3.5ex} $ 15 ^{\circ}$
                    & 15.17 & 0.0 & 56.73
                    & 15.16 & 0.0 & 55.95
                    & 15.23 & 0.0 & 58.46 \\
                    \rule[-1ex]{0pt}{3.5ex} $ 30 ^{\circ}$
                    & 30.17 & 0.0 & 56.73
                    & 30.16 & 0.0 & 55.95
                    & 30.23 & 0.0 & 58.45 \\
                    \rule[-1ex]{0pt}{3.5ex} $ 45 ^{\circ}$
                    & 45.17 & 0.0 & 56.73
                    & 45.16 & 0.0 & 55.95
                    & 45.23 & 0.0 & 58.45 \\
                    \rule[-1ex]{0pt}{3.5ex} $ 60 ^{\circ}$
                    & 60.17 & 0.0 & 56.74
                    & 60.16 & 0.0 & 55.95
                    & 60.23 & 0.0 & 58.44 \\
                    \rule[-1ex]{0pt}{3.5ex} $ 75 ^{\circ}$
                    & 75.17 & 0.0 & 56.74
                    & 75.16 & 0.0 & 55.96
                    & 75.23 & 0.0 & 58.44 \\
                    \rule[-1ex]{0pt}{3.5ex} $ 90 ^{\circ}$
                    & 90.17 & 0.0 & 56.74
                    & 90.16 & 0.0 & 55.96
                    & 90.23 & 0.0 & 58.43 \\
                    \rule[-1ex]{0pt}{3.5ex} $ 105 ^{\circ}$
                    & 105.17 & 0.0 & 56.73
                    & 105.16 & 0.0 & 55.96
                    & 105.23 & 0.0 & 58.44 \\
                    \rule[-1ex]{0pt}{3.5ex} $ 120 ^{\circ}$
                    & 120.17 & 0.0 & 56.73
                    & 120.16 & 0.0 & 55.96
                    & 120.23 & 0.0 & 58.44 \\
                    \rule[-1ex]{0pt}{3.5ex} $ 135 ^{\circ}$
                    & 135.17 & 0.0 & 56.73
                    & 135.16 & 0.0 & 55.96
                    & 135.23 & 0.0 & 58.45 \\
                    \rule[-1ex]{0pt}{3.5ex} $ 150 ^{\circ}$
                    & 150.17 & 0.0 & 56.73
                    & 150.16 & 0.0 & 55.96
                    & 150.23 & 0.0 & 58.46 \\
                    \rule[-1ex]{0pt}{3.5ex} $ 165 ^{\circ}$
                    & 165.17 & 0.0 & 56.73
                    & 165.16 & 0.0 & 55.96
                    & 165.23 & 0.0 & 58.46 \\
                    \rule[-1ex]{0pt}{3.5ex} $ 180 ^{\circ}$
                    & 180.17 & 0.0 & 56.73
                    & 180.16 & 0.0 & 55.95
                    & 180.23 & 0.0 & 58.46 \\
                    \hline
                \end{tabular}
            \end{center}
        \end{subtable}     
\end{table}
\subsubsection{Far-field polarization orientation and relative polarization-angle offset}
Table~\ref{tab:polarization_offset_90GHz} to \ref{tab:polarization_offset_230GHz} list the values of the far-field beam polarization angles $\varphi_{\rm beam}$ and the relative polarization-angle offset $\Delta\varphi_{\rm rel}$, for the three off-axis feedhorn positions with a set of polarization orientations. Using the polarization-angle definition described by Method 2 in Sect.~\ref{subsection:method_polar}, the values of the maximum cross-polarization discrimination of the optical system can also be determined and included in the tables. 

The results show that, for off-axis beams, the far field polarization direction is not exactly aligned with the nominal input feedhorn polarization direction. This offset arises because the optical system is no longer symmetric with respect to the displaced feed position. For example for an $x$-polarized input feedhorn, $\varphi_{\rm feed}=0^\circ$ at Position 3, the resulting far-field beam shows a polarization-angle offset of $0.23^\circ$ with respect to the defined optical reference frame. This demonstrates that the off-axis reflector geometry introduces a small non-zero feed-position-dependent offset of the far-field polarization orientation.

Although the far-field polarization orientation shows a non-zero feed-position-dependent offset with respect to the defined optical reference frame for off-axis beams, the relative polarization-angle offset $\Delta\varphi_{\rm rel}$ remains consistent with zero as $\varphi_{\rm feed}$ is varied from $0^\circ$ to $180^\circ$. This indicates that the reflector-induced offset is mainly a fixed polarization-angle shift for a given feed position, rather than a differential rotation that depends on the input feedhorn polarization orientation. The far-field polarization-orientation offsets for the three representative off-axis feed positions are summarized in Table~\ref{tab:polarization_offset_absolute}.
\begin{figure}[h]
  \centering
  \begin{minipage}[b]{1\textwidth}
    \centering
    \includegraphics[width=\textwidth]{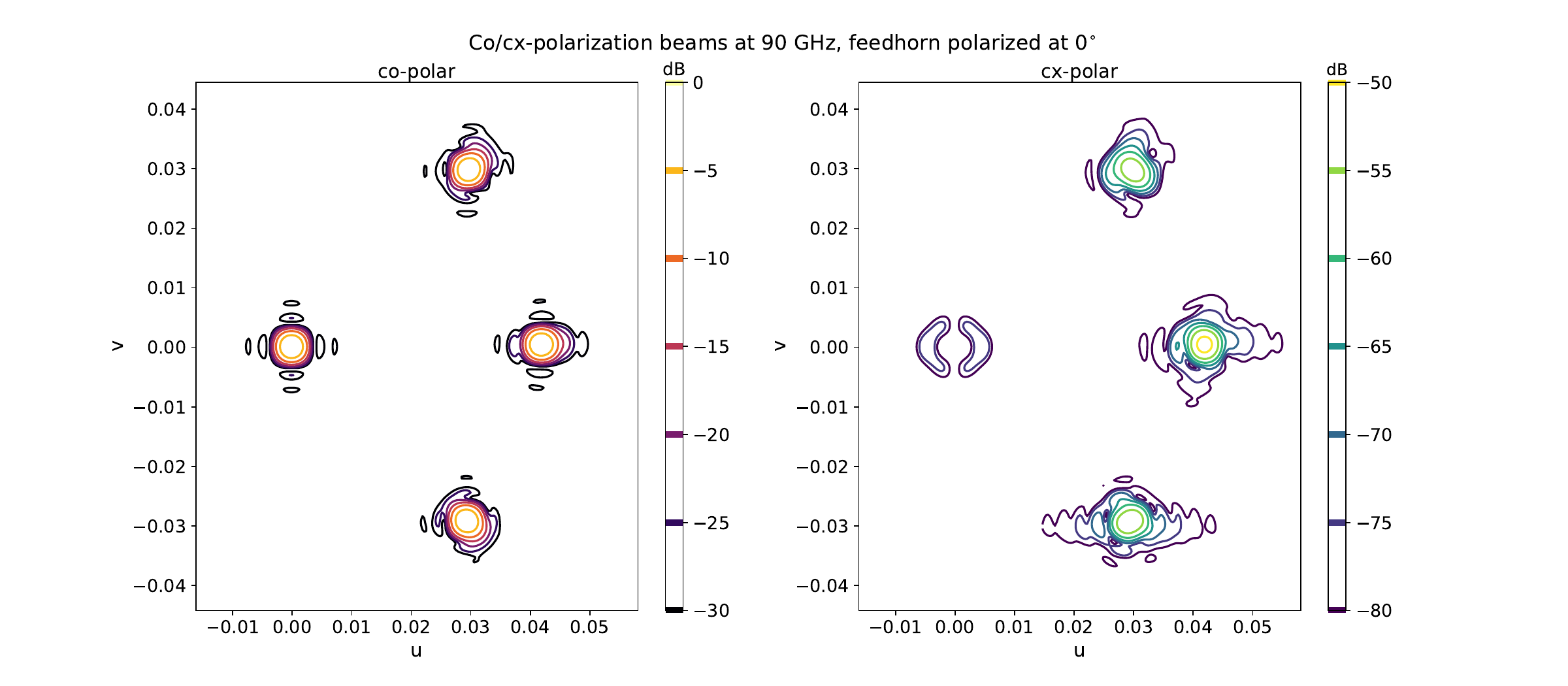}
  \end{minipage}
  \begin{minipage}[b]{1\textwidth}
    \centering
    \includegraphics[width=\textwidth]{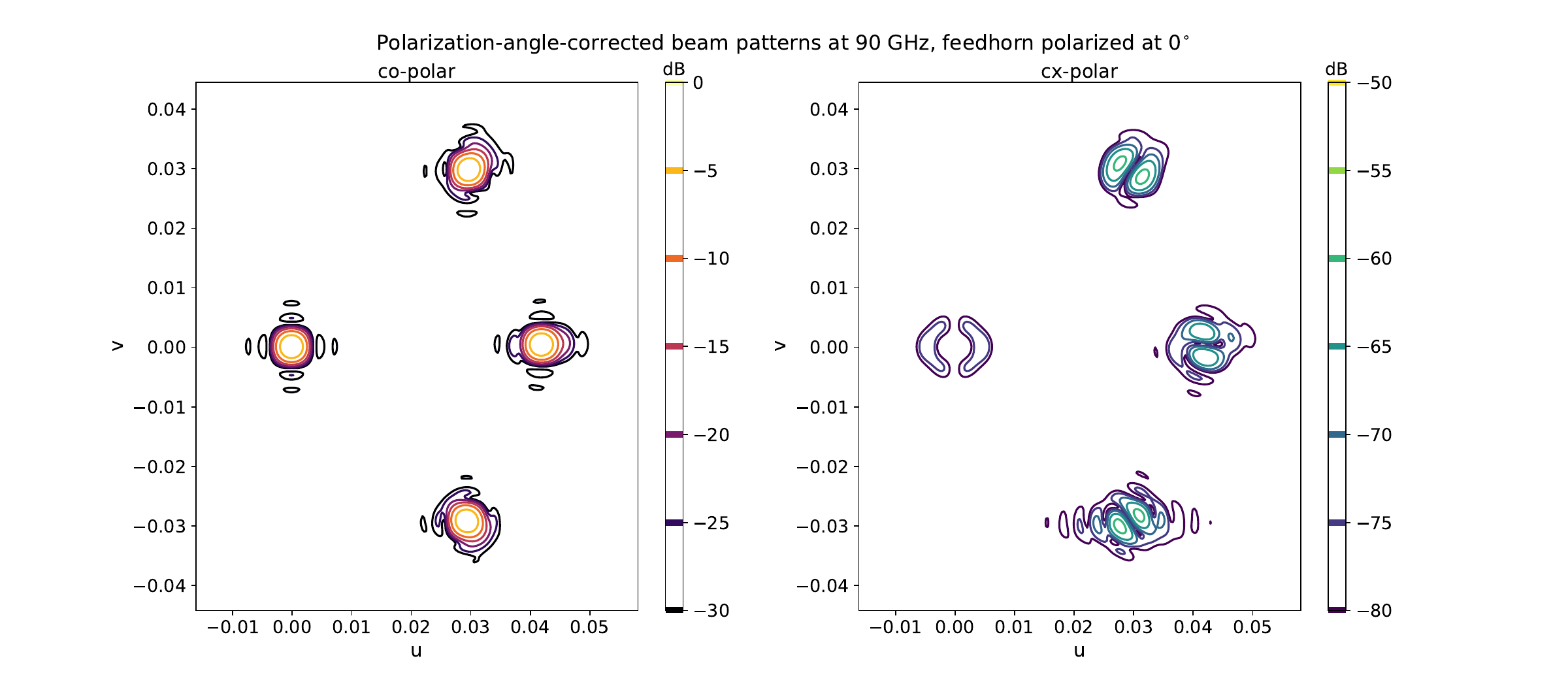}
  \end{minipage}
  \caption{Comparison of the co-polar and cross-polar far-field beam patterns at \SI{90}{\giga\hertz} before and after correcting for the far-field polarization-orientation offset. The upper panels show the original co-polar and cross-polar components evaluated in the nominal feedhorn polarization basis, while the lower panels show the corresponding beam patterns after rotating the polarization basis by the measured polarization-orientation offset. The correction aligns the analysis basis with the actual far-field beam polarization direction and significantly reduces the cross-polarization level. The maps are shown in dB in the $(u,v)$ sky coordinates. For visualization, the beam size shown in this figure is enlarged by a factor of five.}
  \label{fig:far_field_corrections}
\end{figure}
\begin{table}[h]
\caption{Far-field polarization-orientation offset in the $C_{\rm sky}$ coordinate system for a given feedhorn polarization angle defined in $C_{\rm fp}$ } 
\label{tab:polarization_offset_absolute}
        \begin{subtable}[b]{\textwidth}
            \begin{center}  
                \begin{tabular}{|c|c|c|c|}
                    \hline
                     \rule[-1ex]{0pt}{3.5ex} Feedhorn position: 
                     & \multicolumn{1}{c|}{1} 
                     &  \multicolumn{1}{c|}{2} 
                     &  \multicolumn{1}{c|}{3}  \\
                    \hline
                    \rule[-1ex]{0pt}{3.5ex}  Beam polarization-orientation offset: 
                    & $0.17^{\circ}$ &$0.16^{\circ}$ & $0.23^{\circ}$ \\
                    \hline 
                \end{tabular}
            \end{center}
        \end{subtable}     
\end{table}

To illustrate the effect of the beam polarization-orientation offset, Fig.~\ref{fig:far_field_corrections} compares the co-polar and cross-polar components before and after correcting for this offset. The correction is applied by rotating the polarization basis by the measured offset, so that it is aligned with the far-field beam polarization direction. For CMB polarization measurements, this correction is important because an uncorrected beam polarization-angle offset can mix Stokes $Q$ and $U$ and produce leakage between $E$- and $B$-mode signals. The reduction of the cross-polarization level after correction shows that accurately determining the far-field beam polarization direction is necessary for distinguishing true optical cross-polarization from a basis-mismatch effect.

This polarization-angle analysis was also repeated using Method~1 and obtained fully consistent results. To avoid redundancy, the corresponding values are not listed separately. In the simulations, the feedhorn at Position~0 is located on the optical axis and on the symmetry axis of the optical system. Owing to this symmetry, no polarization-angle offset is expected for this feed position. The corresponding far-field beam patterns for Position 0, before and after polarization-angle correction, are shown in Fig.~\ref{fig:far_field_corrections}. The far-field beam for Position~0, center feedhorn, also gives the best cross-polarization discrimination among the simulated feedhorn positions, with a maximum value of approximately \SI{75}{\decibel}.

\subsubsection{Effect of Observing frequency and illumination edge taper}
We also examine the dependence of the polarization response on observing frequency and illumination edge taper. The results show that the far-field beam polarization angle and the relative polarization-angle offset remain unchanged across the simulated frequencies. The maximum cross-polarization discrimination also remains nearly consistent, indicating that the polarization response of the reflector system is not sensitive to observing frequency over the range considered here.

The effect of illumination edge taper was tested using ideal Gaussian feed beams with edge tapers of $-8$, $-12$, and \SI{-20}{\decibel}. The resulting changes in the beam polarization angle are below $0.01^\circ$, and the relative polarization-angle offset remains zero. Therefore, within the ideal Gaussian-feed model used in this work, the polarization-angle response is mainly determined by the feed position in the focal plane and the crossed-Dragone reflector geometry, rather than by the observing frequency or illumination edge taper.

The polarization-angle analysis was also repeated using Method~1 and obtained fully consistent results. To avoid redundancy, the corresponding values are not listed separately. In the simulations, the feedhorn at Position~0 is located on the optical axis and on the symmetry axis of the optical system. Owing to this symmetry, no polarization-angle offset is expected for this feed position.  

\subsection{Polarization response for the tilted-feed configuration}
In the simulations above, the feedhorn was assumed to sit in a flat focal plane, which is the xy-plane in the coordinate frame $C_{\rm fp}$. But in practice, the focal plane the coma-corrected CD optics has a curved focal plane which is spherical surface with the center at 17.5 meters in front of the focal plane \cite{parshley2018optical}. To maintain telecentricity, each feedhorn is tilted so that its symmetry axis follows the local surface normal. The tilted-feedhorn configuration is therefore closer to the actual receiver design of the SO LAT.

\begin{figure}[h]
    \centering
    \includegraphics[width=1\linewidth]{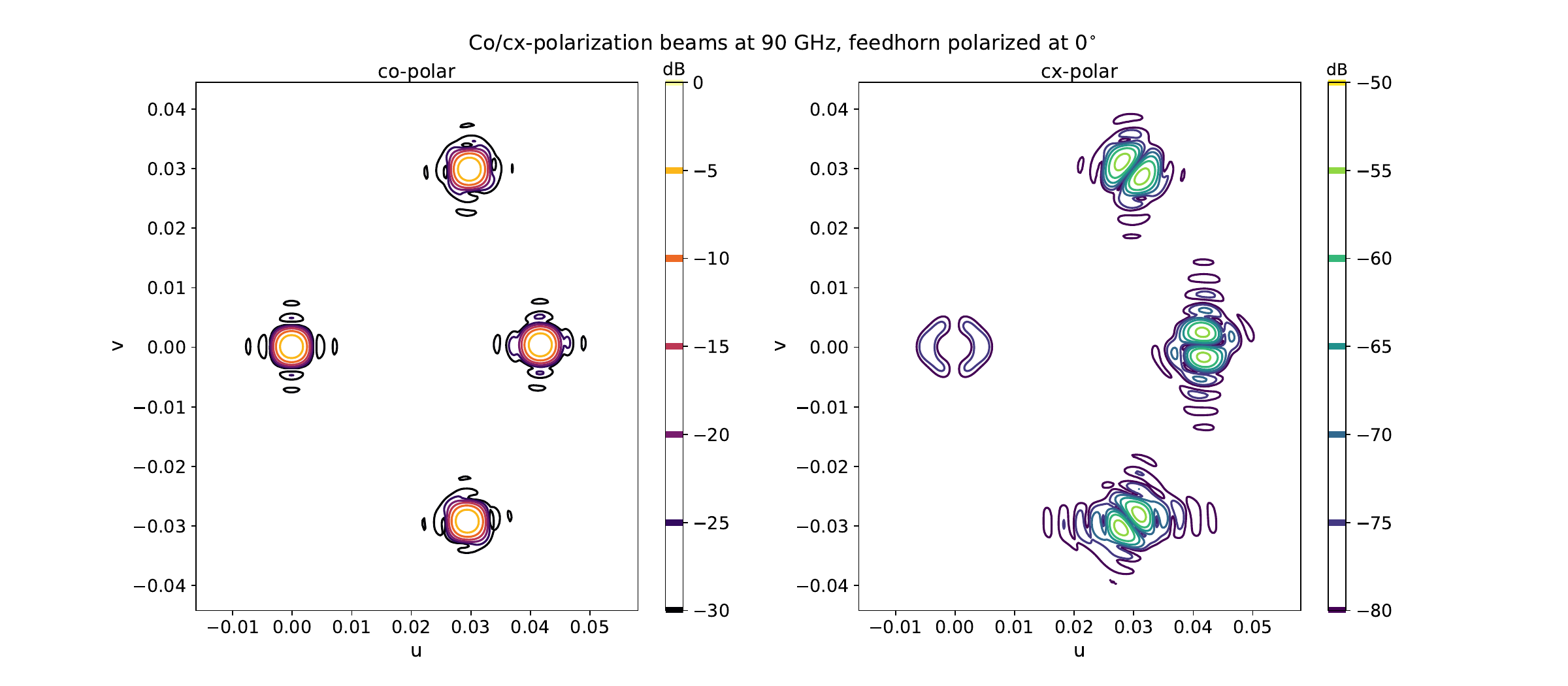}
    \caption{Simulated co- and cross-polar beam patterns at 90 GHz for the tilted-feedhorn configuration. In this configuration, each feedhorn axis is aligned with the local normal direction of the curved focal surface. Compared with the untilted-feedhorn case shown in Fig.~\ref{fig:far_field_corrections}, the off-axis beams become noticeably more symmetric.}
    \label{fig:tilted_Feedhorn_beams}
\end{figure}

Fig~\ref{fig:tilted_Feedhorn_beams} shows the simulated co- and cross-polar beam patterns at 90 GHz for this telecentric curved-focal-plane configuration. Compared with the untilted-feedhorn case shown in Fig.~\ref{fig:far_field_corrections}, the off-axis beams become noticeably more symmetric.

We further evaluate the polarization-angle response in this configuration by rotating each feedhorn about its own symmetry axis. The feedhorn polarization angle, $\varphi_{\rm beam}$, is varied from $0^\circ$ to $180^\circ$ in steps of $15^\circ$, and the corresponding far-field beam polarization angle on the sky is calculated using the same polarization-angle analysis described above. The results show that the relative polarization-angle offset remains consistent with zero for the tilted-feedhorn configuration.

\section{Summary and future work}
In this work, we studied the far-field beam polarization response of the SO LAT reflector system using PO simulations. The simulations were performed at 90, 150, and 230 GHz for three off-axis feed positions, different input feedhorn polarization orientations, and different illumination edge tapers. The far-field beam polarization angle was evaluated using two independent methods: one based on the beam-averaged Stokes parameters and the other based on the maximization of cross-polarization discrimination. The two methods give consistent values for the beam polarization angle, confirming the robustness of the polarization-angle definition used in this work.

For off-axis feed position, the far-field beam polarization direction is not exactly identical to the nominal feedhorn polarization direction, leading to a small polarization-orientation offset in the adopted sky coordinate system. The three simulated feed positions are located about 600 mm from the focal plane center, corresponding to $2.4^{\circ}$ pointing direction from the telescope's boresight, close to the edge of the middle-frequency (MF) optical tube. Among these positions, Position 3 gives the largest beam polarization offset, with a value of $0.23^\circ$. Our simulations also show that the polarization-angle offsets caused by the reflector system are independent of observing frequency and illumination edge taper. These results suggest that the polarization offsets in the far-field beams introduced by the crossed-Dragone reflector optics for the MF and high-frequency (HF) modules of the SO LAT are expected to be at or below the level of $0.23^\circ$, or at least of the same order. 

At the same time, the relative polarization-angle offset is found to be consistent with zero within the numerical uncertainty. This indicates that the SO LAT reflector system preserves the relative orientation of different input feedhorn polarization states. This is beneficial because the reflector-induced effect can be treated as a common polarization-basis rotation and does not introduce additional complexity in the reconstruction of $Q$ and $U$. Nevertheless, further work is  required to examine whether the beam patterns remain consistent for different feedhorn polarization orientations, beyond the polarization-angle analysis presented here.

One limitation of the present analysis is that the feedhorn beam is modeled as an ideal circular Gaussian beam with no intrinsic cross-polarization component. This assumption allows us to isolate the polarization response introduced by the SO LAT reflector system itself. In practice, however, the actual feedhorn beam may deviate from a perfect circular Gaussian profile, and its intrinsic cross-polarization response may not be negligible. They will be investigated in future work using realistic feedhorn beam models.

The present simulations also do not include the refractive optical tubes of the SO LAT. In the full SO LAT optical system, the MF and HF modules use a three-lens refractive design, and these additional optical elements can further modify the beam polarization response. In particular, transmission through the lenses and their anti-reflection coatings may affect the $s$- and $p$-polarized components differently, leading to additional cross-polarization response and possible polarization-angle rotation. The anti-reflection coating effects have been discussed in our paper \cite{Ren:26}, where we also introduced a hybrid-PO analysis method that allows the influence of anti-reflection coatings to be included in physical-optics simulations. We will apply this method to the refractive optics of the SO LAT and SAT in future beam simulations and polarization-response analyses.

These results show that the SO LAT reflector system has a stable and well-behaved polarization response, confirming that the reflector design is well suited for CMB polarization measurements. We hope that the reflector-system simulations and analysis presented in this work will provide useful input for the polarization-angle calibration of the SO LAT and for future beam-systematics analyses.

\acknowledgments  
Funded in part by the European Union (ERC, CMBeam,
101040169). JEG gratefully acknowledges support from the University of Iceland Research Fund and the Icelandic Research Fund (Grant number: 2410656-051). We thank Steve Choi, Mark Devlin, Simon Dicker, and Patricio Gallardo for helpful discussions. 

\bibliography{report} 

@article{Minami2020,
  title = {New Extraction of the Cosmic Birefringence from the Planck 2018 Polarization Data},
  author = {Minami, Yuto and Komatsu, Eiichiro},
  journal = {Phys. Rev. Lett.},
  volume = {125},
  issue = {22},
  pages = {221301},
  numpages = {6},
  year = {2020},
  month = {Nov},
  publisher = {American Physical Society},
  doi = {10.1103/PhysRevLett.125.221301},
  url = {https://link.aps.org/doi/10.1103/PhysRevLett.125.221301}
}

@article{SOScience_2019,
doi = {10.1088/1475-7516/2019/02/056},
url = {https://doi.org/10.1088/1475-7516/2019/02/056},
year = {2019},
month = {feb},
publisher = {},
volume = {2019},
number = {02},
pages = {056},
author = {Ade, Peter and Aguirre, James and Ahmed, Zeeshan and Aiola, Simone and Ali, Aamir and Alonso, David and Alvarez, Marcelo A. and Arnold, Kam and Ashton, Peter and Austermann, Jason and Awan, Humna and Baccigalupi, Carlo and Baildon, Taylor and Barron, Darcy and Battaglia, Nick and Battye, Richard and Baxter, Eric and Bazarko, Andrew and Beall, James A. and Bean, Rachel and Beck, Dominic and Beckman, Shawn and Beringue, Benjamin and Bianchini, Federico and Boada, Steven and Boettger, David and Bond, J. Richard and Borrill, Julian and Brown, Michael L. and Bruno, Sarah Marie and Bryan, Sean and Calabrese, Erminia and Calafut, Victoria and Calisse, Paolo and Carron, Julien and Challinor, Anthony and Chesmore, Grace and Chinone, Yuji and Chluba, Jens and Cho, Hsiao-Mei Sherry and Choi, Steve and Coppi, Gabriele and Cothard, Nicholas F. and Coughlin, Kevin and Crichton, Devin and Crowley, Kevin D. and Crowley, Kevin T. and Cukierman, Ari and D'Ewart, John M. and Dünner, Rolando and de Haan, Tijmen and Devlin, Mark and Dicker, Simon and Didier, Joy and Dobbs, Matt and Dober, Bradley and Duell, Cody J. and Duff, Shannon and Duivenvoorden, Adri and Dunkley, Jo and Dusatko, John and Errard, Josquin and Fabbian, Giulio and Feeney, Stephen and Ferraro, Simone and Fluxà, Pedro and Freese, Katherine and Frisch, Josef C. and Frolov, Andrei and Fuller, George and Fuzia, Brittany and Galitzki, Nicholas and Gallardo, Patricio A. and Ghersi, Jose Tomas Galvez and Gao, Jiansong and Gawiser, Eric and Gerbino, Martina and Gluscevic, Vera and Goeckner-Wald, Neil and Golec, Joseph and Gordon, Sam and Gralla, Megan and Green, Daniel and Grigorian, Arpi and Groh, John and Groppi, Chris and Guan, Yilun and Gudmundsson, Jon E. and Han, Dongwon and Hargrave, Peter and Hasegawa, Masaya and Hasselfield, Matthew and Hattori, Makoto and Haynes, Victor and Hazumi, Masashi and He, Yizhou and Healy, Erin and Henderson, Shawn W. and Hervias-Caimapo, Carlos and Hill, Charles A. and Hill, J. Colin and Hilton, Gene and Hilton, Matt and Hincks, Adam D. and Hinshaw, Gary and Hložek, Renée and Ho, Shirley and Ho, Shuay-Pwu Patty and Howe, Logan and Huang, Zhiqi and Hubmayr, Johannes and Huffenberger, Kevin and Hughes, John P. and Ijjas, Anna and Ikape, Margaret and Irwin, Kent and Jaffe, Andrew H. and Jain, Bhuvnesh and Jeong, Oliver and Kaneko, Daisuke and Karpel, Ethan D. and Katayama, Nobuhiko and Keating, Brian and Kernasovskiy, Sarah S. and Keskitalo, Reijo and Kisner, Theodore and Kiuchi, Kenji and Klein, Jeff and Knowles, Kenda and Koopman, Brian and Kosowsky, Arthur and Krachmalnicoff, Nicoletta and Kuenstner, Stephen E. and Kuo, Chao-Lin and Kusaka, Akito and Lashner, Jacob and Lee, Adrian and Lee, Eunseong and Leon, David and Leung, Jason S.-Y. and Lewis, Antony and Li, Yaqiong and Li, Zack and Limon, Michele and Linder, Eric and Lopez-Caraballo, Carlos and Louis, Thibaut and Lowry, Lindsay and Lungu, Marius and Madhavacheril, Mathew and Mak, Daisy and Maldonado, Felipe and Mani, Hamdi and Mates, Ben and Matsuda, Frederick and Maurin, Loïc and Mauskopf, Phil and May, Andrew and McCallum, Nialh and McKenney, Chris and McMahon, Jeff and Meerburg, P. Daniel and Meyers, Joel and Miller, Amber and Mirmelstein, Mark and Moodley, Kavilan and Munchmeyer, Moritz and Munson, Charles and Naess, Sigurd and Nati, Federico and Navaroli, Martin and Newburgh, Laura and Nguyen, Ho Nam and Niemack, Michael and Nishino, Haruki and Orlowski-Scherer, John and Page, Lyman and Partridge, Bruce and Peloton, Julien and Perrotta, Francesca and Piccirillo, Lucio and Pisano, Giampaolo and Poletti, Davide and Puddu, Roberto and Puglisi, Giuseppe and Raum, Chris and Reichardt, Christian L. and Remazeilles, Mathieu and Rephaeli, Yoel and Riechers, Dominik and Rojas, Felipe and Roy, Anirban and Sadeh, Sharon and Sakurai, Yuki and Salatino, Maria and Rao, Mayuri Sathyanarayana and Schaan, Emmanuel and Schmittfull, Marcel and Sehgal, Neelima and Seibert, Joseph and Seljak, Uros and Sherwin, Blake and Shimon, Meir and Sierra, Carlos and Sievers, Jonathan and Sikhosana, Precious and Silva-Feaver, Maximiliano and Simon, Sara M. and Sinclair, Adrian and Siritanasak, Praween and Smith, Kendrick and Smith, Stephen R. and Spergel, David and Staggs, Suzanne T. and Stein, George and Stevens, Jason R. and Stompor, Radek and Suzuki, Aritoki and Tajima, Osamu and Takakura, Satoru and Teply, Grant and Thomas, Daniel B. and Thorne, Ben and Thornton, Robert and Trac, Hy and Tsai, Calvin and Tucker, Carole and Ullom, Joel and Vagnozzi, Sunny and Engelen, Alexander van and Lanen, Jeff Van and Winkle, Daniel D. Van and Vavagiakis, Eve M. and Vergès, Clara and Vissers, Michael and Wagoner, Kasey and Walker, Samantha and Ward, Jon and Westbrook, Ben and Whitehorn, Nathan and Williams, Jason and Williams, Joel and Wollack, Edward J. and Xu, Zhilei and Yu, Byeonghee and Yu, Cyndia and Zago, Fernando and Zhang, Hezi and Zhu, Ningfeng and The Simons Observatory collaboration},
title = {The Simons Observatory: science goals and forecasts},
journal = {Journal of Cosmology and Astroparticle Physics},
}

@article{o2007systematic,
  title={Systematic errors in cosmic microwave background polarization measurements},
  author={O'Dea, Daniel and Challinor, Anthony and Johnson, Bradley R},
  journal={Monthly Notices of the Royal Astronomical Society},
  volume={376},
  number={4},
  pages={1767--1783},
  year={2007},
  publisher={The Royal Astronomical Society}
}

@article{PhysRevD.77.083003,
  title = {CMB polarization systematics due to beam asymmetry: Impact on inflationary science},
  author = {Shimon, Meir and Keating, Brian and Ponthieu, Nicolas and Hivon, Eric},
  journal = {Phys. Rev. D},
  volume = {77},
  issue = {8},
  pages = {083003},
  numpages = {11},
  year = {2008},
  month = {Apr},
  publisher = {American Physical Society},
  doi = {10.1103/PhysRevD.77.083003},
  url = {https://link.aps.org/doi/10.1103/PhysRevD.77.083003}
}

@article{franco2003systematic,
  title={Systematic effects in the measurement of polarization by the PLANCK telescope},
  author={Franco, G and Fosalba, Pablo and Tauber, Jan A},
  journal={Astronomy \& Astrophysics},
  volume={405},
  number={1},
  pages={349--366},
  year={2003},
  publisher={EDP Sciences}
}

@inproceedings{cornelison2022improved,
  title={Improved polarization calibration of the BICEP3 CMB polarimeter at the South Pole},
  author={Cornelison, James and Verges, Clara and Ade, PAR and Ahmed, Zeeshan and Amiri, Mandana and Barkats, Denis and Thakur, R Basu and Beck, Dominic and Bischoff, Colin A and Bock, James J and others},
  booktitle={Millimeter, Submillimeter, and Far-Infrared Detectors and Instrumentation for Astronomy XI},
  volume={12190},
  pages={829--848},
  year={2022},
  organization={SPIE}
}

@inproceedings{parshley2018optical,
  title={The optical design of the six-meter CCAT-prime and Simons Observatory telescopes},
  author={Parshley, Stephen C and Niemack, Michael and Hills, Richard and Dicker, Simon R and D{\"u}nner, Rolando and Erler, Jens and Gallardo, Patricio A and Gudmundsson, Jon E and Herter, Terry and Koopman, Brian J and others},
  booktitle={Ground-based and Airborne Telescopes VII},
  volume={10700},
  pages={1292--1304},
  year={2018},
  organization={SPIE}
}

@inproceedings{parshley2018ccat,
  title={CCAT-prime: a novel telescope for sub-millimeter astronomy},
  author={Parshley, Stephen C and Kronshage, J{\"o}rg and Blair, James and Herter, Terry and Nolta, Mike and Stacey, Gordon J and Bazarko, Andrew and Bertoldi, Frank and Bustos, Ricardo and Campbell, Donald B and others},
  booktitle={Ground-based and Airborne Telescopes VII},
  volume={10700},
  pages={1744--1758},
  year={2018},
  organization={SPIE}
}

@article{Zhu_2021,
doi = {10.3847/1538-4365/ac0db7},
url = {https://doi.org/10.3847/1538-4365/ac0db7},
year = {2021},
month = {sep},
publisher = {The American Astronomical Society},
volume = {256},
number = {1},
pages = {23},
author = {Zhu, Ningfeng and Bhandarkar, Tanay and Coppi, Gabriele and Kofman, Anna M. and Orlowski-Scherer, John L. and Xu, Zhilei and Adachi, Shunsuke and Ade, Peter and Aiola, Simone and Austermann, Jason and Bazarko, Andrew O. and Beall, James A. and Bhimani, Sanah and Bond, J. Richard and Chesmore, Grace E. and Choi, Steve K. and Connors, Jake and Cothard, Nicholas F. and Devlin, Mark and Dicker, Simon and Dober, Bradley and Duell, Cody J. and Duff, Shannon M. and Dünner, Rolando and Fabbian, Giulio and Galitzki, Nicholas and Gallardo, Patricio A. and Golec, Joseph E. and Haridas, Saianeesh K. and Harrington, Kathleen and Healy, Erin and Ho, Shuay-Pwu Patty and Huber, Zachary B. and Hubmayr, Johannes and Iuliano, Jeffrey and Johnson, Bradley R. and Keating, Brian and Kiuchi, Kenji and Koopman, Brian J. and Lashner, Jack and Lee, Adrian T. and Li, Yaqiong and Limon, Michele and Link, Michael and Lucas, Tammy J and McCarrick, Heather and Moore, Jenna and Nati, Federico and Newburgh, Laura B. and Niemack, Michael D. and Pierpaoli, Elena and Randall, Michael J. and Sarmiento, Karen Perez and Saunders, Lauren J. and Seibert, Joseph and Sierra, Carlos and Sonka, Rita and Spisak, Jacob and Sutariya, Shreya and Tajima, Osamu and Teply, Grant P. and Thornton, Robert J. and Tsan, Tran and Tucker, Carole and Ullom, Joel and Vavagiakis, Eve M. and Vissers, Michael R. and Walker, Samantha and Westbrook, Benjamin and Wollack, Edward J. and Zannoni, Mario},
title = {The Simons Observatory Large Aperture Telescope Receiver},
journal = {The Astrophysical Journal Supplement Series}
}

@inproceedings{xd_holo_FYST,
author = {Xiaodong Ren and Pablo Astudillo and Urs U. Graf and Richard E. Hills and Sebastian Jorquera and Bojan Nikolic and Stephen C. Parshley and Nicol{\'a}s Reyes and Lars Weikert},
title = {{Holographic surface measurement system for the Fred Young Submillimeter Telescope}},
volume = {11445},
booktitle = {Ground-based and Airborne Telescopes VIII},
editor = {Heather K. Marshall and Jason Spyromilio and Tomonori Usuda},
organization = {International Society for Optics and Photonics},
publisher = {SPIE},
pages = {114456D},
year = {2020},
doi = {10.1117/12.2560459},
URL = {https://doi.org/10.1117/12.2560459}
}

@article{Ren:26,
author = {Xiaodong Ren and Rustam Balafendiev and Jon E. Gudmundsson},
journal = {Appl. Opt.},
number = {16},
pages = {5659--5666},
publisher = {Optica Publishing Group},
title = {Polarization offset through differential transmission in refractive CMB telescopes identified using a hybrid physical optics method},
volume = {65},
month = {Jun},
year = {2026},
url = {https://opg.optica.org/ao/abstract.cfm?URI=ao-65-16-5659},
doi = {10.1364/AO.591740},
}

@book{collin1969antenna,
  title={Antenna theory: PART 1},
  author={Collin, Robert E and Zucker, Francis J},
  year={1969},
  publisher={McGraw-Hill Inc.,US},
  
}

@book{ufimtsev2014fundamentals,
  title={Fundamentals of the physical theory of diffraction},
  author={Ufimtsev, Pyotr Ya},
  year={2014},
  publisher={John Wiley \& Sons}
}

@ARTICLE{Xu2021,
       author = {{Xu}, Zhilei and {Adachi}, Shunsuke and {Ade}, Peter and {Beall}, J.~A. and {Bhandarkar}, Tanay and {Bond}, J. Richard and {Chesmore}, Grace E. and {Chinone}, Yuji and {Choi}, Steve K. and {Connors}, Jake A. and {Coppi}, Gabriele and {Cothard}, Nicholas F. and {Crowley}, Kevin D. and {Devlin}, Mark and {Dicker}, Simon and {Dober}, Bradley and {Duff}, Shannon M. and {Galitzki}, Nicholas and {Gallardo}, Patricio A. and {Golec}, Joseph E. and {Gudmundsson}, Jon E. and {Haridas}, Saianeesh K. and {Harrington}, Kathleen and {Hervias-Caimapo}, Carlos and {Patty Ho}, Shuay-Pwu and {Huber}, Zachary B. and {Hubmayr}, Johannes and {Iuliano}, Jeffrey and {Kaneko}, Daisuke and {Kofman}, Anna M. and {Koopman}, Brian J. and {Lashner}, Jack and {Limon}, Michele and {Link}, Michael J. and {Lucas}, Tammy J. and {Matsuda}, Frederick and {McCarrick}, Heather and {Nati}, Federico and {Niemack}, Michael D. and {Orlowski-Scherer}, John and {Piccirillo}, Lucio and {Sarmiento}, Karen Perez and {Schaan}, Emmanuel and {Silva-Feaver}, Maximiliano and {Sonka}, Rita and {Sutariya}, Shreya and {Tajima}, Osamu and {Teply}, Grant P. and {Terasaki}, Tomoki and {Thornton}, Robert and {Tucker}, Carole and {Ullom}, Joel and {Vavagiakis}, Eve M. and {Vissers}, Michael R. and {Walker}, Samantha and {Whipps}, Zachary and {Wollack}, Edward J. and {Zannoni}, Mario and {Zhu}, Ningfeng and {Zonca}, Andrea and {Simons Observatory Collaboration}},
        title = "{The Simons Observatory: The Large Aperture Telescope (LAT)}",
      journal = {Research Notes of the American Astronomical Society},
         year = 2021,
        month = apr,
       volume = {5},
       number = {4},
          eid = {100},
        pages = {100},
          doi = {10.3847/2515-5172/abf9ab},
archivePrefix = {arXiv},
       eprint = {2104.09511},
 primaryClass = {astro-ph.IM},
       adsurl = {https://ui.adsabs.harvard.edu/abs/2021RNAAS...5..100X}
}

@article{Gudmundsson:21,
author = {Jon E. Gudmundsson and Patricio A. Gallardo and Roberto Puddu and Simon R. Dicker and Alexandre E. Adler and Aamir M. Ali and Andrew Bazarko and Grace E. Chesmore and Gabriele Coppi and Nicholas F. Cothard and Nadia Dachlythra and Mark Devlin and Rolando D\"{u}nner and Giulio Fabbian and Nicholas Galitzki and Joseph E. Golec and Shuay-Pwu Patty Ho and Peter C. Hargrave and Anna M. Kofman and Adrian T. Lee and Michele Limon and Frederick T. Matsuda and Philip D. Mauskopf and Kavilan Moodley and Federico Nati and Michael D. Niemack and John Orlowski-Scherer and Lyman A. Page and Bruce Partridge and Giuseppe Puglisi and Christian L. Reichardt and Carlos E. Sierra and Sara M. Simon and Grant P. Teply and Carole Tucker and Edward J. Wollack and Zhilei Xu and Ningfeng Zhu},
journal = {Appl. Opt.},
number = {4},
pages = {823--837},
publisher = {Optica Publishing Group},
title = {The Simons Observatory: modeling optical systematics in the Large Aperture Telescope},
volume = {60},
month = {Feb},
year = {2021},
url = {https://opg.optica.org/ao/abstract.cfm?URI=ao-60-4-823},
doi = {10.1364/AO.411533},
}

@inproceedings{Dicker2018,
author = {S. R. Dicker and P. A. Gallardo and J. E. Gudmundsson and P. D. Mauskopf and A. Ali and P. C. Ashton and G. Coppi and M. J. Devlin and N. Galitzki and S. P. Ho and C. A. Hill and J. Hubmayr and B. Keating and A. T. Lee and M. Limon and F. Matsuda and J. McMahon and M. D. Niemack and J. L. Orlowski-Scherer and L. Piccirillo and M. Salatino and S. M. Simon and S. T. Staggs and R. Thornton and J. N. Ullom and E. M. Vavagiakis and E. J. Wollack and Z. Xu and N. Zhu},
title = {{Cold optical design for the large aperture Simons Observatory telescope}},
volume = {10700},
booktitle = {Ground-based and Airborne Telescopes VII},
editor = {Heather K. Marshall and Jason Spyromilio},
organization = {International Society for Optics and Photonics},
publisher = {SPIE},
pages = {107003E},
year = {2018},
doi = {10.1117/12.2313444},
URL = {https://doi.org/10.1117/12.2313444}
}

@INPROCEEDINGS{Gallardo2018,
       author = {{Gallardo}, Patricio A. and {Gudmundsson}, Jon and {Koopman}, Brian J. and {Matsuda}, Frederick T. and {Simon}, Sara M. and {Ali}, Aamir and {Bryan}, Sean and {Chinone}, Yuji and {Coppi}, Gabriele and {Cothard}, Nicholas and {Devlin}, Mark J. and {Dicker}, Simon and {Fabbian}, Giulio and {Galitzki}, Nicholas and {Hill}, Charles A. and {Keating}, Brian and {Kusaka}, Akito and {Lashner}, Jacob and {Lee}, Adrian T. and {Limon}, Michele and {Mauskopf}, Philip D. and {McMahon}, Jeff and {Nati}, Federico and {Niemack}, Michael D. and {Orlowski-Scherer}, John L. and {Parshley}, Stephen C. and {Puglisi}, Giuseppe and {Reichardt}, Christian L. and {Salatino}, Maria and {Staggs}, Suzanne and {Suzuki}, Aritoki and {Vavagiakis}, Eve M. and {Wollack}, Edward J. and {Xu}, Zhilei and {Zhu}, Ningfeng},
        title = "{Systematic uncertainties in the Simons Observatory: optical effects and sensitivity considerations}",
    booktitle = {Millimeter, Submillimeter, and Far-Infrared Detectors and Instrumentation for Astronomy IX},
         year = 2018,
       editor = {{Zmuidzinas}, Jonas and {Gao}, Jian-Rong},
       series = {Society of Photo-Optical Instrumentation Engineers (SPIE) Conference Series},
       volume = {10708},
        month = jul,
          eid = {107083Y},
        pages = {107083Y},
          doi = {10.1117/12.2312971},
archivePrefix = {arXiv},
       eprint = {1808.05152},
 primaryClass = {astro-ph.IM},
       adsurl = {https://ui.adsabs.harvard.edu/abs/2018SPIE10708E..3YG}
}

@article{ludwig1973,
title={The definition of cross polarization},
author={Ludwig, Arthur},
journal={IEEE Transactions on Antennas and Propagation},
volume={21},
number={1},
pages={116--119},
year={1973},
publisher={IEEE}
}

@article{2025ApJS27934B,
doi = {10.3847/1538-4365/ade0bd},
url = {https://doi.org/10.3847/1538-4365/ade0bd},
year = {2025},
month = {jul},
publisher = {The American Astronomical Society},
volume = {279},
number = {2},
pages = {34},
author = {Bhandarkar, Tanay and Haridas, Saianeesh K. and Iuliano, Jeff and Kofman, Anna and Manduca, Alex and Perez Sarmiento, Karen and Orlowski-Scherer, John and Satterthwaite, Thomas P. and Wang, Yuhan and Ahmed, Zeeshan and Austermann, Jason E. and Bae, Kyuyoung and Coppi, Gabriele and Devlin, Mark J. and Dicker, Simon R and Dow, Peter N. and Duff, Shannon M. and Dutcher, Daniel and Galitzki, Nicholas and Gudmundsson, Jon E. and Henderson, Shawn W. and Hubmayr, Johannes and Johnson, Bradley R. and Koc, Matthew A. and Koopman, Brian J. and Limon, Michele and Link, Michael J and Lucas, Tammy J. and Moore, Jenna E. and Nati, Federico and Niemack, Michael D. and Sierra, Carlos E. and Silva-Feaver, Max and Singh, Robinjeet and Sonka, Rita F. and Staggs, Suzanne T. and Thornton, Robert J. and Tsan, Tran and Van Lanen, Jeff L. and Vavagiakis, Eve M. and Vissers, Michael R and Walters, Liam and Zannoni, Mario and Zheng, Kaiwen},
title = {Simons Observatory: Characterization of the Large Aperture Telescope Receiver},
journal = {The Astrophysical Journal Supplement Series}
}

@article{Murphy:24,
author = {Colin C. Murphy and Steve K. Choi and Rahul Datta and Mark J. Devlin and Matthew Hasselfield and Brian J. Koopman and Jeff McMahon and Sigurd Naess and Michael D. Niemack and Lyman A. Page and Suzanne T. Staggs and Robert Thornton and Edward J. Wollack},
journal = {Appl. Opt.},
number = {19},
pages = {5079--5087},
publisher = {Optica Publishing Group},
title = {Optical modeling of systematic uncertainties in detector polarization angles for the Atacama Cosmology Telescope},
volume = {63},
month = {Jul},
year = {2024},
url = {https://opg.optica.org/ao/abstract.cfm?URI=ao-63-19-5079},
doi = {10.1364/AO.521079},
}
\bibliographystyle{spiebib} 
\end{document}